\documentclass[]{aa}
\usepackage{graphicx,longtable,lscape}
\usepackage{txfonts,psfig}
\usepackage[]{natbib}
\def \ferg {erg cm$^{-2}$ s$^{-1}$}

\def\ltsima{$\; \buildrel < \over \sim \;$}
\def\lsim{\lower.5ex\hbox{\ltsima}}
\def\gtsima{$\; \buildrel > \over \sim \;$}
\def\gsim{\lower.5ex\hbox{\gtsima}}

\newcommand{\be}{\begin{equation}}
\newcommand{\en}{\end{equation}}

\def\msole {~M_{\odot}}

\begin{document}
  \title{Swift-XRT 6-year monitoring of the ultraluminous X-ray source M33-X8}
\author{ V.\ La Parola\inst{1}, A. D'A\'i\inst{1}, G. Cusumano\inst{1}, T. Mineo\inst{1}}

   \offprints{V. La Parola, laparola@ifc.inaf.it}
   \institute{$^1$~INAF -- Istituto di Astrofisica Spaziale e Fisica Cosmica di Palermo,
        Via U.\ La Malfa 153, 90146 Palermo, Italy  
 }

\abstract
{The long term evolution of ULX with their spectral
and luminosity variations in time give important clues on the nature 
of ULX and on the accretion process that powers them. }
{We report here the results of a Swift-XRT 6-year monitoring campaign of  the 
closest example of a persistent ULX, M33 X-8, that extends to 16 years the monitoring of this source in
the soft X-rays.  The luminosity of this source is a few $10^{39}$ erg 
s$^{-1}$, marking the faint end of the ULX luminosity function. }
{We  analysed  the set of 15 observations collected during the Swift monitoring. 
We searched for differences in the spectral parameters at different observing 
epochs, adopting several models commonly used to fit the X-ray spectra of ULX.}
{The source
exhibits flux variations of the order of 30\%. No significant spectral
variations are observed along the monitoring. The average 0.5-10 keV spectrum can be well
described by a thermal model, either in the form of a slim disk, or as a
combination of a Comptonized corona and a standard accretion disk.}
{}
\keywords{X-rays: general  - X-rays: individuals: M33 X-8 }
\authorrunning {V. La Parola  et al.}
\titlerunning {Swift monitoring of M33 X-8}

\maketitle

\section{Introduction\label{intro} }

Ultraluminous X-ray sources are point-like, off-nuclear objects observed in many
nearby galaxies to have isotropic luminosity between $\sim 10^{39}$ and 
$\sim 10^{41}$ erg $\rm s^{-1}$ (e.g. \citealp{fabbiano89,swartz11}). There are 
several hypotheses to explain their nature (and indeed they may form an heterogenous class of
sources): if the emission is isotropic, then it exceeds the Eddington limit for a
stellar mass black hole, and could indicate the presence of an intermediate
mass black hole (IMBH, with $\rm M_{BH}\sim 100 - 1000 \msole$, e.g.
\citealp{colbert99,sutton12}), whose existence may be related either to Population III
stars \citep{madau01,fryer01}, or to the capture and stripping of the nuclei of 
satellite galaxies in hierarchical merging \citep{king01b}, or to repeated mergers
of stellar mass black holes in globular clusters \citep{miller02}. On the 
other hand, the emission could be either 
relativistically beamed (for example, \citealp{begelman06} investigate on the 
analogy of the Galactic microquasar SS433 with the ULX class), or, more likely, 
geometrically beamed (i.e. collimated into a wind-produced funnel, see, e.g., 
\citealp{king01,king09}), or 
we could be seeing a super-Eddington ultraluminous accretion state
(\citealp{gladstone09} and reference therein): all these  mechanisms would allow for
more common stellar mass black holes (with $\rm M_{BH}\lesssim 100 \msole$). The
recent discovery of a 1.37s pulsation in the ULX M82-X2 \citep{bachetti14} has set the
case for the presence of neutron stars in the ULX population \citep{king09}, 
triggering a renewed interest in this yet challenging debate.

A strong X-ray emission (which is persistent in most cases, although there 
are also some remarkable example of transient ULXs, see e.g., 
\citealp{middleton12,soria15}) is ubiquitous to these sources, 
whereas only a few of them are detected at other wavelenghts. Thus, the main tools to 
gain knowledge on their nature are the analysis of their X-ray spectra to
identify the main physical processes that power them, and the study of their 
light curves to understand how these processes are correlated with each other 
and with the luminosity of the sources. Several studies, based both on samples of 
ULXs (e.g. \citealp{gladstone09, stobbart06}) and on the monitoring of single 
sources (e.g. \citealp{kong10,feng10, grise10}) have been carried on 
in this direction, showing that in most cases the emission can be described 
with a combination of a thermal disk-like component, plus a (broken) power 
law-like component (see also \citealp{feng11} for a review). The 
relative contribution of the two components (if both are present) as 
well as their temperature/slope 
may vary substantially from source to source.  The observed
phenomenology, that presents several evident inconsitencies with that of Galactic
Black Hole (GBH) binaries (e.g., the persistence in  
a bright state of most ULX with smooth spectral variations, as opposite to the 
transient behaviour of accreting GBHs, the frequent presence of a soft thermal 
excess below 2 keV, a spectral curvature at $\sim 3-5$ keV, see e.g.,
\citealp{gladstone09,soria11,vierdayanti10,middleton15}) have been combined in a model that 
describes the ULXs
as accreting black holes whose emission is powered by supercritical accretion. In
this model the disk appears as a standard one \citep{shakura73} at large 
radii, and emerges as a slim disk in the inner region, providing a moderate 
super-Eddington luminosity \citep{abramowicz88,watarai00, ebisawa03}. 
The accreting mass in excess of the critical Eddington limit may be ejected 
through a collimated wind, resulting in a geometrical beaming. This wind
is transparent at small radii, and leaves the 
innermost (hot) region of the disk exposed to the viewer 
\citep{poutanen07,middleton15}. At larger viewing angles, the optically
thick wind hides the innermost regions of the disk
resulting in an apparently lower inner temperature.
A wide spectrum of black hole masses and/or 
different viewing angles (as discussed, e.g. in
\citealp{middleton15,sutton13}) may thus explain the observed variegated 
phenomenology. 

\begin{figure}
\begin{center}
\centerline{\includegraphics[width=12cm,angle=0]{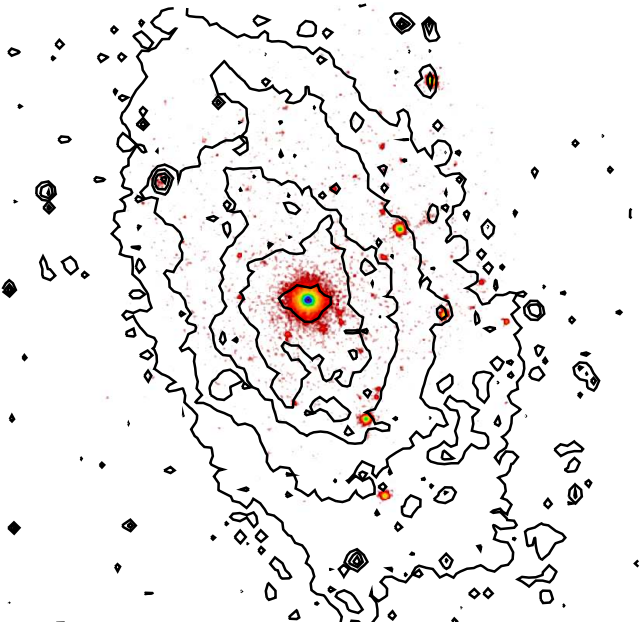}}
\caption[]{X-ray image of M33 obtained
cumulating all the Swift-XRT observations, with superimposed the 
optical contour levels, showing the position of M33 X-8 coincident with the
optical center of the galaxy. Contour levels are derived from the DSS 
image (http://archive.eso.org/dss/dss) and have a linear spacing in
intensity with a constant ratio of 0.13 beetween consecutive levels and the
faintest level corresponding to 10\% of the maximum intensity in the image. 

}
                \label{img} 
        \end{center}
        \end{figure}

M33 X-8 \citep{long81,trinchieri88}, located at a distance of $\sim 820$ kpc 
\citep{freedman01} is the closest example of persistent ULX. Its position is roughly
coincident with the center of its host galaxy (Figure~\ref{img}), but the upper limit derived by
\citet{gebhardt01} on the mass of the nucleus of M33 ($1500 \msole$) and the
detection of a $\sim 106$ d periodical modulation \citep{dubus97} rules out 
the possibility that the source is a low luminosity AGN. On the other hand no
optical counterpart could be identified, due to the extreme crowding of the field.
Its X-ray luminosity of a few
$\rm 10^{39} erg~s^{-1}$ makes it belong to the faint end of the ULX
luminosity distribution. Its flux, persistently above 
$\rm 10^{-11} erg~cm^{-2} s^{-1}$ in the X-ray band, allows very detailed
studies of its spectrum and how it evolves in time (see 
\citealp{foschini06,weng09,middleton11,isobe12}). 

In this paper we present the results of an observing campaign on M33 X-8 performed with 
Swift-XRT \citep{swift}. The paper is organized as follows. Section 2 describes
the data and their reduction; section 3 reports on the results of the spectral 
analysis; in section 4 we discuss the results and draw our conclusions.

\section{Observations and data reduction\label{data}}
Swift-XRT \citep{xrt} observed the central region of M33 fifteen times between 
December 2007 and June 2013, with two different campaign, targeted to M33 X-8 and to
Nova2010-10a respectively. In the latter group of observations M33 X-8 is 
$\sim 5.5$ arcmin off-axis. The details on all the observations are reported in 
table~\ref{log}. Figure~\ref{img} shows the XRT image of the source
obtained after integrating over all the observations, with a total exposure time of 
$\sim 115$ ksec.
All the observations are in Photon Counting observing mode \citep{hill04}. 
The data were processed with standard procedures ({\sc xrtpipeline}) 
using the {\sc ftools} in the {\sc heasoft} package (v 6.16) and the 
products were extracted adopting a grade filtering of 0-12. The 
source count rate in all the observations varies in a range where we may 
expect some photon pile-up. Therefore, for the spectral analysis, we checked 
each observation for the presence of pile-up by comparing the
source radial profile with the expected PSF profile \citep{moretti05} and 
excluding the inner region where the two curves diverge 
\footnote{see {\tt http://www.swift.ac.uk/analysis/xrt/pileup.php} for a 
complete description of this procedure}. In observation \#10 the source is crossed
by a hot column through its centroid, so this observation was not used for the 
spectral analysis.
The background was extracted for all the observations from a 50-pixel radius 
circular region far from other bright point sources in the field. 
The ancillary response files for each spectrum were generated with 
{\sc xrtmkarf} and we used the spectral redistribution matrix v013
\footnote{http://heasarc.gsfc.nasa.gov/docs/heasarc/caldb/swift}.
We also built the average spectrum summing the spectra from the single
observations (using {\sc mathpha}); the relevant ancillary files were 
combined using {\sc addarf}, weighting them according to the exposure 
time of the corresponding spectra.

To allow the use of $\chi^2$ statistics all the spectra were rebinned to
have at least 20 counts per energy bin.  The spectral analysis was performed 
using {\sc xspec} v.12.5.

\begin{table*}
\begin{center}
\begin{tabular}{r l l r r r}
\hline
Obs \# &Obs ID      & Date       & Elapsed Time & Exposure  & Flux$_{0.3-10 keV}$\\ 
       &            &            &  (ks)        & (ks)      & $10^{-11}$\ferg\\ \hline
1      &00031042001 & 2007-12-26 & 12.564       & 2.944     &  $1.44\pm 0.08 $\\
2      &00031856001 & 2010-11-03 & 40.533       & 5.054     &  $1.51\pm 0.06 $\\
3      &00031856002 & 2010-11-07 & 18.752       & 6.011     &  $1.65\pm 0.06 $\\
4      &00031856003 & 2010-11-11 & 47.384       & 5.946     &  $1.55\pm 0.06 $\\
5      &00031856004 & 2010-11-15 & 29.981       & 6.068     &  $1.52\pm 0.06 $\\
6      &00031856005 & 2010-11-23 & 46.673       & 5.586     &  $1.60\pm 0.06 $\\
7      &00031856006 & 2010-12-01 & 47.244       & 6.166     &  $1.46\pm 0.05 $\\
8      &00031856007 & 2010-12-09 & 33.849       & 5.944     &  $1.54\pm 0.06 $\\
9      &00031856008 & 2010-12-18 & 19.002       & 6.580     &  $1.60\pm 0.06 $\\
10     &00031856009 & 2010-12-25 & 13.003       & 3.207     &  --\\
11     &00031856010 & 2011-01-02 & 46.342       & 6.213     &  $1.60\pm 0.07 $\\
12     &00031042002 & 2012-11-05 & 64.329       & 19.591    &  $1.82\pm 0.05 $\\
13     &00031042003 & 2013-02-06 & 69.409       & 18.730    &  $1.44\pm 0.04 $\\
14     &00031042004 & 2013-06-10 & 64.253       & 14.179    &  $1.73\pm 0.04 $\\
15     &00031042005 & 2013-06-13 & 23.980       & 5.661     &  $1.37\pm 0.06 $\\
\hline
\end{tabular}
\caption{XRT observations log. ObsID 00031856009 was discarded because of the
presence of a hot column crossing the source centroid. The last column reports
the observed flux derived using the best fit {\sc diskpbb} model. \label{log}}
\end{center}
\end{table*}

\section{Analysis and results\label{an.spec} }

As a first step, we fit simultaneously  the fourteen spectra obtained from 
the single observations, constraining 
the model parameters to assume the same value for 
all the spectra, except for the model normalization (parametrized through a 
multiplicative constant fixed to 1 for the faintest spectrum, i.e. Obs \# 15,
and left free to vary for the others).

We tested two single component models: a power-law,
that has been used to describe  ULXs in their hard state (see e.g. 
\citealp{winter06}; however, see also \citealp{gladstone09,bachetti13,walton14}, 
that illustrates the limits of this model in the presence of high statistics 
data), and a modified disk model ({\sc diskpbb} in {\sc xspec}),
i.e. an accretion disk model where a parameter (p) describes the temperature 
radial dependance as $\rm T\propto R^{-p}$: a value of 0.75 indicate a 
standard disk, expected for sources in a pure thermal state 
\citep{makishima00,winter06}, while a value p=0.5 describes a slim disk, where 
advective energy transport dominates over radiative cooling (see, e.g., 
\citealp{watarai00}). An absorption  component ({\sc phabs}, with abundances
from \citealp{anders89}) was
included in each of the test models, with absorbing column fixed to the 
Galactic line-of-sight value of $\rm 1.1\times10^{21} cm^{-2}$ \citep{kalberla05}. 
A second absorption component was included as a free parameter to describe 
any intrinsic absorption, if present  (i.e. the absorption in each model is
described as {\sc phabs}$_{\rm Gal}\times${\sc phabs}$_{\rm local}$, where the
former is fixed to the Galactic line of sight value and the latter is a free fit
parameter for all the tested models).

\begin{figure}
\begin{center}
\centerline{\includegraphics[width=9cm,angle=0]{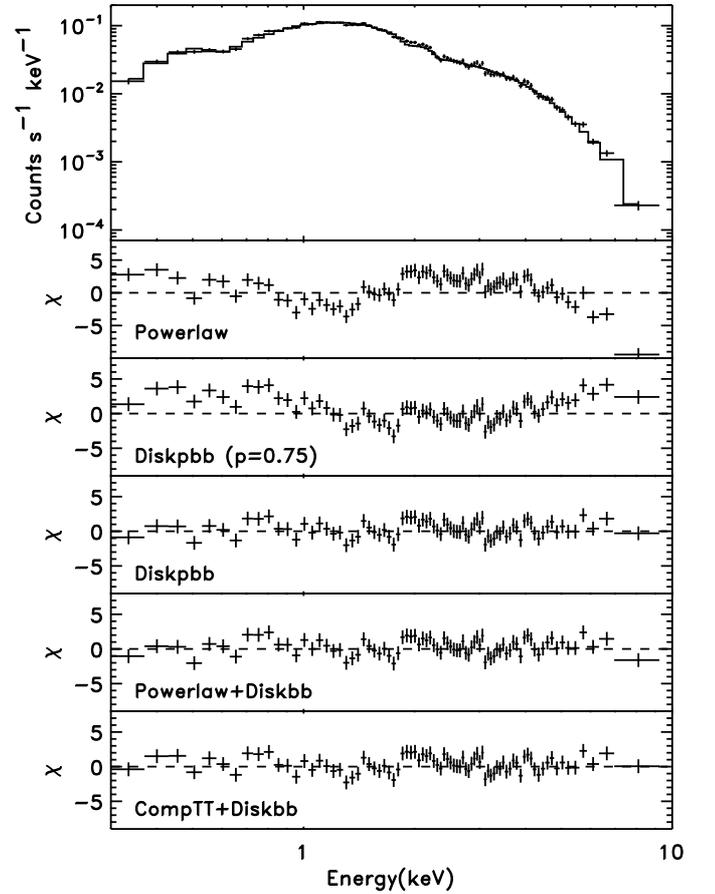}}
\caption[]{Top panel: data and best fit model (with model {\sc diskpbb}) for 
the average M33 X-8 spectrum. Lower panels: residuals for the different 
spectral models reported in table~\ref{spfit2}
}
                \label{spectrum} 
        \end{center}
        \end{figure}

The power law model yielded a poor fit,
with apparent trends in the residuals in all the datasets (see also the 
average spectrum residuals in
Figure~\ref{spectrum}, panel b). Even after releasing
the power law indices and the intrinsic absorbing column, allowing them to vary 
independently for each data set, the fit did not generally improve, showing 
no significant change in the residuals pattern, with  $\chi^2=1649.7/1457$ 
d.o.f.  

The {\sc diskpbb} model was first tested leaving the inner disk temperature as a free
parameter with a common value for all the datasets, while the  temperature 
radial dependence was constrained to be the one expected
for a standard accretion disk, fixing the parameter p to 0.75. This setting 
is not adequate, as it produces a best fit model that clearly underestimates 
the source emission at both ends of the data energy range (see also the 
average spectrum residuals in Figure~\ref{spectrum}, panel c). These 
sistematics do
persist when we release the temperature for each single spectrum.

\begin{figure}
\begin{center}
\centerline{\includegraphics[width=9cm,angle=0]{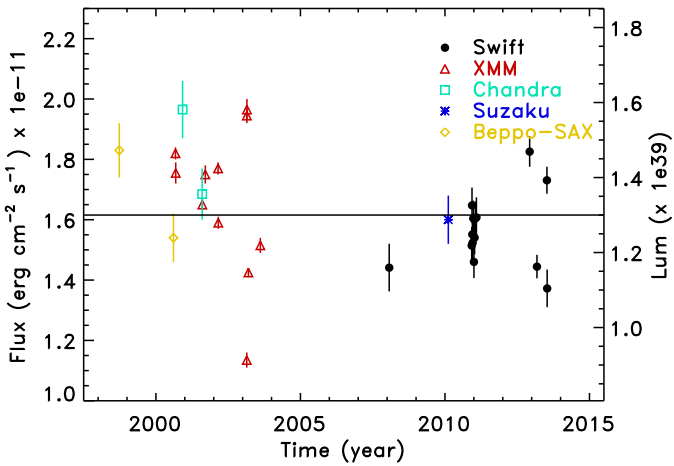}}
\caption[]{Long term light curve of M33 X-8. Each point 
corresponds to a single observation, and fluxes are not corrected for 
absorption. The luminosity on the right axis has been evaluated
assuming a distance of 820 kpc. We have associated an arbitrary 5\% statistical error 
to the BeppoSAX, Chandra and Suzaku points. The horizontal line represents the
Eddington luminosity for a $\rm 10 M_{\odot}$ black hole. 
}
                \label{lc} 
        \end{center}
        \end{figure}

Instead, we obtain a  good fit, with no significant trends in the residuals 
(and $\rm \chi^2/dof=1452.3/1483$), when we leave both p and the 
temperatureas free fit parameters, with 
common values for all the datasets.  The best fit parameters for 
this fit are $\rm T_{in}=1.48\pm0.08$ keV, $\rm p=0.60\pm0.02$, and 
$\rm N_{Hi}=0.031\pm 0.015$ (here and in the following the uncertainties 
reported for each spectral parameters are at 90\,\% confidence level). We have
also verified that letting the model parameters vary independently for each data 
set do not improve the fit significantly, with paramenters values consistent
within their errors among the single spectra. The 
observed (i.e. not corrected for absorption)
fluxes resulting for each  spectrum from this best 
fit model (evaluated using the {\sc cflux} convolution model in 
{\sc xspec}) are reported in Table~\ref{log} and plotted in Figure~\ref{lc}
(black circles).
The source has an average 0.3-10 keV absorbed flux of 
$(1.57\pm 0.02) \times10^{-11}$ \ferg,
and variations of $\lesssim15\%$ around this value. The average luminosity, 
assuming isotropic emission at the distance on 817 Mpc \citep{freedman01} is 
$\sim1.3 \times10^{39}$ erg
s$^{-1}$.  

As the source does not show significant spectral variation throughout our 
monitoring, we used the average spectrum obtained from the entire dataset to
refine the spectral analysis.
Table~\ref{spfit2} reports the best fit spectral parameters obtained for the
average spectrum using the {\sc diskpbb} model. For completeness, we report also 
the fit results and residuals obtained with a
simple power law model and with a standard accretion disk model (p=0.75): both
fits are not statistically acceptable.
In all models, an additional absorbing column was included to verify the 
presence of any intrinsic absorption. The 
absorbing column reported in the table is the best fit value in excess of the 
Galactic line of sight value. 
Figure~\ref{spectrum} shows the data with the best fit model and the residuals 
relevant to the different models reported in Table~\ref{spfit2}. 
We have also tested two two-component models, both used to describe a
disk+corona geometry. 
In particular, we used a model composed by a 
power law plus a multicolor disk  ({\sc diskbb}), that results in 
a hot disk (the inner temperature is 1.15 keV) and a soft power law 
(with photon index 2.13), and a {\sc diskbbb+comptt} model (with the seed
photons temperature tied to the disk peak temperature), whose best fit 
parameters
suggest a cool disk ($\sim 0.50$ keV) and a cool (with an electron temperature
of $\sim 1.1$ keV) and optically thick ($\tau\sim 14$) corona. Both models provided
statistically acceptable fits for our data.

\begin{figure}
\begin{center}
\centerline{\includegraphics[width=9cm,angle=0]{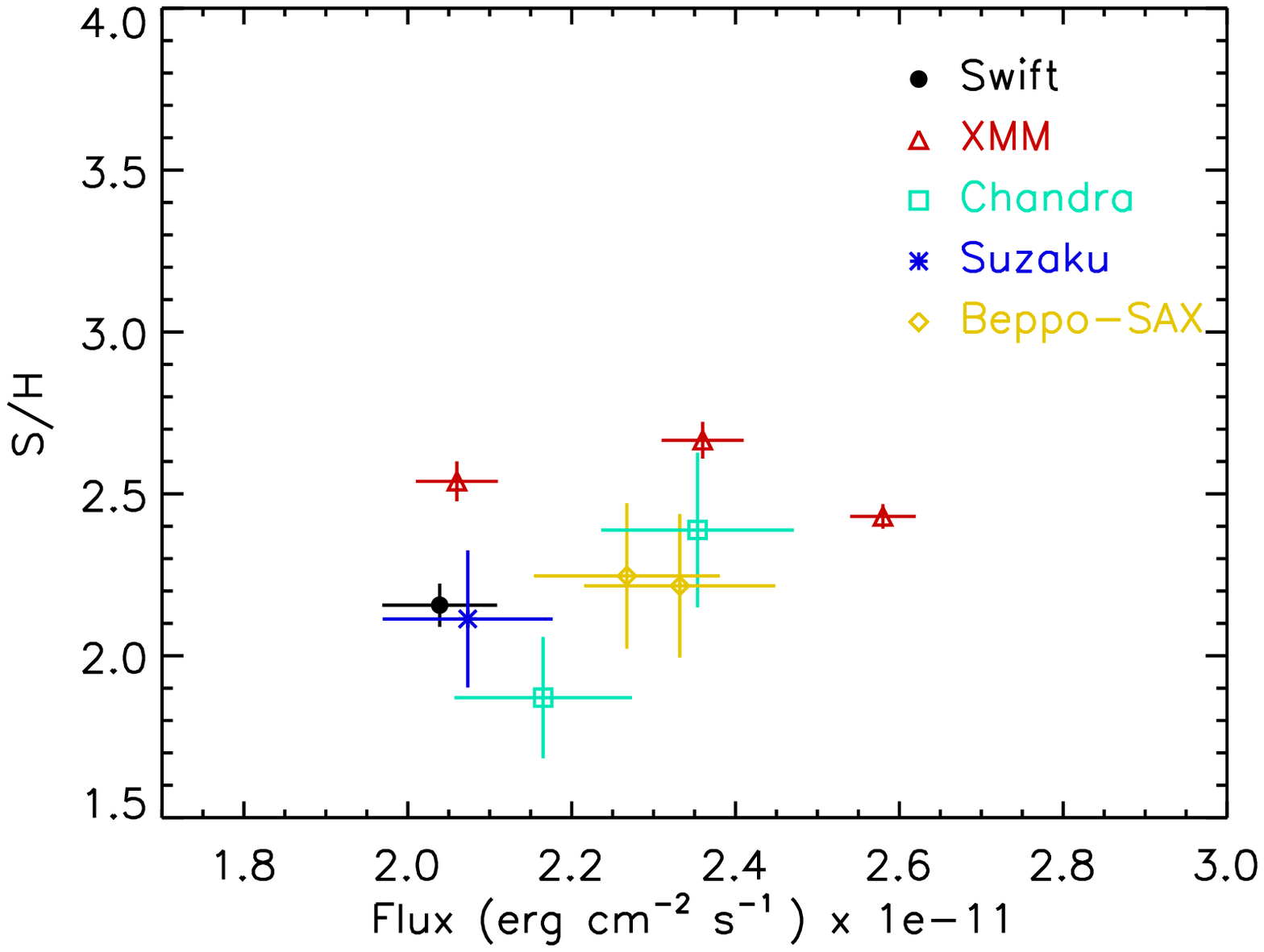}}
\caption[]{{Spectral colors evaluated as the ratio of the fluxes in the 0.3-3 keV (S) 
and 3-10 keV (H) bands. For the other datasets we used the best fit models reported by
\citet{parmar01} (Beppo-SAX), \citet{laparola03,dubus04} (Chandra),
\citet{middleton11} (XMM-Newton),
and \citet{isobe12} (Suzaku). The error on the ratio has been evaluated assuming the fractional 
error on each band to be equal to the fractional error on the total flux, when
reported, or assuming an arbitrary error of 5\% on the flux in each band
when the error on the flux was missing. All fluxes are corrected for 
absorption.}

}
                \label{sat} 
        \end{center}
        \end{figure}

Finally, we have compared our spectral results with those derived in the past with
other satellites.  To this aim we have reported in Figure~\ref{sat} the spectral 
colors derived from the best fit model for the data analysed in this
work and for all the datasets where enough spectral information is available in 
literature. The plot shows that the spectral colors (evaluated as the ratio 
of the fluxes in the 0.3-3 keV and 3-10 keV bands) are mostly insensitive 
to the flux variations, showing that the spectral shape is not significantly 
variable within the current statistics used to constrain the spectrum.

\begin{table*}
\begin{center}
\begin{tabular}{l|r l r r r } \hline
Model           & $\rm N_{Hi} (\times 10^{22})$ &  Parameters                 & Flux $(\times 10^{-11})$   &$\chi^2$/dof  \\
                & cm$^{-2}$           &                             & \ferg                              &              \\ \hline \hline
Power law       &$\rm 0.17 \pm 0.01$  &$\Gamma=2.18_{-0.03}^{+0.03}$        &$1.95_{-0.03}^{+0.03}$      &   882.2/476  \\ \hline
Diskpbb         &$ 0$                 &kT= $1.10_{-0.17}^{+0.17}$ keV       &$1.77_{-0.02}^{+0.02}$      &   678.4/476  \\
(p=0.75)        &                     &Rcos$\theta=62_{-2}^{+2}$ km         &                            &              \\ \hline
Diskpbb         &$0.045 \pm 0.016$    &kT= $1.43_{-0.07}^{+0.07}$ keV       &$2.04_{-0.07}^{+0.08}$      &   518.2/475  \\
                &                     &Rcos$\theta=27_{-4}^{+4}$  km        &                            &              \\
		&                     &p=$0.60_{-0.02}^{+0.02}$             &                            &              \\ \hline
Powerlaw+diskbb &$0.06 \pm 0.04$      &$\Gamma=2.1_{-0.3}^{+0.3}$           &$1.0_{-0.3}^{+0.2}$         &   521.2/474  \\
                &                     &kT=$1.12_{-0.08}^{+0.09}$  keV       &$1.16_{-0.10}^{+0.10}$      &              \\
		&                     &Rcos$\theta=49_{-7}^{+9}$  km        &                            &              \\ \hline
Diskbb+CompTT   &$\sim0$              &$\rm kT_{disk}=0.58_{-0.11}^{+0.21}$ &$0.9_{-0.2}^{+0.4}$         &   513.4/473  \\
                &                     &Rcos$\theta=160_{-50}^{+60}$         &                            &              \\
                &                     &$\rm kT_{p}=1.27_{-0.11}^{+0.17}$    &$1.0_{-0.4}^{+0.2}$         &              \\
                &                     &$\tau=11_{-2}^{+7}$                  &                            &              \\
\hline
\end{tabular}
\caption{Averaged spectrum of the 14 Swift/XRT observations: 
best fit results.  N$_{Hi}$ is the absorbing column in excess of the Galactic
value. For each spectral component we report the intrinsic flux in the 0.3-10
keV range. \label{spfit2}}
\end{center}
\end{table*}



\section{Discussion\label{discussion}}
We have investigated the spectral properties of the ultraluminous X-ray
source M33 X-8, the closest persistent source of its class  (820 Mpc), 
located in the vicinity of the nucleus of the nearby galaxy M33, through a 
Swift-XRT monitoring, that consists of 15 observations, spanning 6 years. 

M33 X-8 shows a weak flux variability over the entire XRT monitoring. 
Figure~\ref{lc} shows the 0.3--10 keV 16-year long term light 
curve of M33 X-8, reporting also the 0.3-10 keV flux (not corrected for 
absorption) observed in the past by SAX \citep{parmar01}, Chandra 
\citep{laparola03,dubus04}, XMM-Newton \citep{middleton11}, and Suzaku \citep{isobe12}.
The flux variations observed with Swift are consistent with what 
observed with the other satellites. The luminosity varies between 1.0 and 
1.6$\times 10^{39}$ erg s$^{-1}$, and locates it at the low luminosity end of 
the known ULX sample. Significant long-term flux variability is commonly observed in 
ULX, but the variability amplitude observed in M33 X-8 is 
lower than that observed in other persistent ULXs, that may reach a factor of
$\sim 5$ in flux amplitude, as shown, e.g., by Ho IX X-1, that shows such wide 
flux variations on a monthly scale \citep{laparola01,vierdayanti10}, but several
other examples can be found, e.g, in the sample analysed  by \citet{pintore14}.

The energy spectrum obtained from the averaged Swift-XRT spectra shows an apparent
curvature, that makes it largely inconsistent with a simple power law. Instead,
it can be well described by a thermal model: in particular, we obtained a very
good description using either a disk model with a modified temperature profile 
\citep{watarai00}, or the two component models.

The simplest two component model (power law + disk) is a phenomenological 
model often  used to describe the spectra of ULXs as an empirical
description of a disk plus corona geometry. In the presence of a cool
(kT$\sim 0.1-0.4$ keV) and luminous (L$\sim 10^{39}-10^{40}$ erg/s) disk, it 
allows to infer the presence of intermediate mass black holes
(e.g, \citealp{makishima00}). This is not the case for M33 X-8, where the disk
component describes well the high energy part of the spectrum, and
appears hot (kT$\sim 1.15$keV), leaving a soft excess that is accounted for by
the powerlaw. The overall disk parameters are then 
inconsistent with a massive black
hole, but instead are more typical of an ordinary stellar mass black hole:  
using the relationship between mass, temperature and luminosity in a standard 
disk (see e.g. \citealp{makishima00}), we derive a mass of $\sim 10
M_{\odot}$ for a non rotating black hole, consistent with the estimation
obtained by data from other satellites (e.g. 
\citealp{foschini06,weng09,isobe12}).
\citet{sutton13} developed a 
classification scheme based on a disk+power law fit, to be applied
to ULX spectra, according to which the spectral state of an ULX source can be
defined by the disk temperature, the power law slope, and the ratio between the
flux contribution of the two spectral component in the 0.3-1 keV band. 
Our result is consistent with that found by \citet{sutton13} using 
XMM-Newton data,
and, according to their classification, it identifies M33 X-8 as a broadened 
disk source, i.e. a source whose spectrum
is dominated by emission from a hot disk (see Table~\ref{spfit2}) and where 
the additional soft component
may be the effect of a poorly realistic description of the disk spectrum by the
{\sc diskbb} model. In fact, such hot disk/soft power law spectra are difficult
to explain in the context of the analogy of ULXs with Galactic black hole
binaries: the thermal state of GBHs is indeed characterized by a hot 
disk, but the presence of a soft powerlaw-like component in addition to the disk is 
unusual, and its physical interpretation is not simple: if this component is
due to the presence of a Comptonized corona, we do not expect it to be 
dominant at energies lower than the temperature of the seed photons, that come
from the disk.  

The same scenario of a Comptonized corona over an accretion disk 
can be modelled in a more physical way with a combination of a
disk spectrum plus a Comptonized spectrum ({\sc diskbb + comptt} in
our modeling). \citet{gladstone09} derive a distinctive spectral sequence from 
the comparison of several sources, including M33 X-8, interpreting it with the
progressive emergence of a (wind driven) corona. A similar interpretation is given
by \citet{middleton12} and \citet{soria15} for M31 ULX-1 and M83 ULX-1, 
respectively, showing how different spectral states correlate with the source
luminosity. In this respect, our results suggest the presence of cool disk plus an 
optically thick corona. In general,  this 
kind of spectrum breaks the similarity between ULXs and Galactic black hole
binaries, where a Comptonized corona over the disk is observed to be hotter
(kT$\gtrsim50$ keV) and thinner ($\tau\lesssim1$) (e.g. \citealp{kubota04}). 
A noteworthy exception among GBHs is the microquasar GRS 1915+105, that, during 
its soft phase (which is associated to near-Eddington accretion), was observed 
to show a low-temperature high-opacity Comptonized spectrum \citep{ueda09}.
Recent simulations \citep{ohsuga09} have shown that the strong radiation
pressure that results from a high accretion rate may induce important outflows 
from the inner part of the disk, resulting in a low temperature, optically 
thick Comptonizing wind 
that blocks the view to the inner and hottest part of the disk. This becomes visible 
again as the radiation pressure decreases and the wind weakens, reducing its 
launching radius.
The accretion flow in faint ULXs such as M33 X-8 may be different
from the one described by these simulations that assume higher accretion rates, 
nevertheless radiation pressure driven winds may be at work in this case as 
well. This mechanism (described for example in
\citealp{middleton12} to explain the spectral behaviour of an ULX in M31) also 
explains the deviation 
of the  luminosity/temperature relation from the one expected form a standard disk 
($\rm L \propto T^4$), as higher luminosities correspond to the disk being 
truncated at larger radii by the radiation pressure of the wind.

The slim disk hypothesis has been already proposed as a physically consistent 
description of the spectrum from this source by 
\citet{weng09}, on the basis of an observing campaign carried on with 
XMM-Newton and by \citet{isobe12},
from the analysis of a Suzaku observation. \citet{middleton11} also suggest
the emergence of a thick Comptonized corona over the disk as the flux 
increases: namely, such a
hard component is required to describe the data of their highest flux bin 
(F$_{0.3-10}>1.9\times10^{-11}$\ferg). With the present data, however, we are 
not able to verify this hypothesis because the source never reaches such a 
high flux level during our monitoring.

If the slim disk interpretation is correct, we find 
a value of the temperature gradient p of $0.60\pm 0.02$,  
i.e. not consistent with the standard disk value of 0.75, thus implying that
the disk is in an advective regime, with a super-critical accretion rate: 
according to \citet{watarai00}, for a mass of $\sim 10 M_{\odot}$, the 
observed luminosity and temperature are consistent with a mass accretion rate 
of a factor of 10 higher than the critical rate. 


\begin{acknowledgements}
This work has been supported by ASI grant I/011/07/0.
\end{acknowledgements}

\bibliographystyle{aa}

\begin{thebibliography}{}

\bibitem[Abramowicz et al.(1988)]{abramowicz88} Abramowicz, M.~A., 
Czerny, B., Lasota, J.~P., \& Szuszkiewicz, E.\ 1988, \apj, 332, 646 

\bibitem[Anders \& Grevesse(1989)]{anders89} Anders, E., \& Grevesse, N.\ 1989,
\gca, 53, 197

\bibitem[Bachetti et al.(2013)]{bachetti13} Bachetti, M., Rana, 
V., Walton, D.~J., et al.\ 2013, \apj, 778, 163 

\bibitem[Bachetti et al.(2014)]{bachetti14} Bachetti, M., 
Harrison, F.~A., Walton, D.~J., et al.\ 2014, \nat, 514, 202 

\bibitem[Begelman et al.(2006)]{begelman06} Begelman, M.~C., King, 
A.~R., \& Pringle, J.~E.\ 2006, \mnras, 370, 399 

\bibitem[Burrows et al.(2004)]{xrt} Burrows, D.~N., Hill, 
J.~E., Nousek, J.~A., et al.\ 2004, \procspie, 5165, 201 

\bibitem[Colbert \& Mushotzky(1999)]{colbert99} Colbert, E.~J.~M., \& Mushotzky, 
R.~F.\ 1999, \apj, 519, 89 

\bibitem[Dubus et al.(2004)]{dubus04} Dubus, G., Charles, P. A., Long, K. S.,
\ 2004, \aap, 425, 95 

\bibitem[Dubus et al.(1997)]{dubus97} Dubus, G., Charles, 
P.~A., Long, K.~S., \& Hakala, P.~J.\ 1997, \apjl, 490, L47 

\bibitem[Ebisawa et al.(2003)]{ebisawa03} Ebisawa, K., {\.Z}ycki, 
P., Kubota, A., Mizuno, T., \& Watarai, K.-y.\ 2003, \apj, 597, 780 

\bibitem[Fabbiano(1989)]{fabbiano89} Fabbiano, G.\ 1989, \araa, 27, 87 

\bibitem[Feng \& Kaaret(2010)]{feng10} Feng, H., \& Kaaret, P.\ 2010, \apjl, 
712, L169 

\bibitem[Feng \& Soria(2011)]{feng11} Feng, H., \& Soria, R.\ 2011, \nar, 55, 166 

\bibitem[Foschini et al.(2006)]{foschini06} Foschini, L., Ebisawa, 
K., Kawaguchi, T., et al.\ 2006, Advances in Space Research, 38, 1378 


\bibitem[Freedman et al.(2001)]{freedman01} Freedman, W.~L., 
Madore, B.~F., Gibson, B.~K., et al.\ 2001, \apj, 553, 47 

\bibitem[Fryer et al.(2001)]{fryer01} Fryer, C.~L., Woosley, 
S.~E., \& Heger, A.\ 2001, \apj, 550, 372 

\bibitem[Gebhardt et al.(2001)]{gebhardt01} Gebhardt, K., Lauer, 
T.~R., Kormendy, J., et al.\ 2001, \aj, 122, 2469 

\bibitem[Gehrels et al. (2004)]{swift} Gehrels, N., et al.\ 
2004, \apj, 611, 1005 

\bibitem[Gladstone et al.(2009)]{gladstone09} Gladstone, J.~C., 
Roberts, T.~P., \& Done, C.\ 2009, \mnras, 397, 1836 

\bibitem[Gris{\'e} et al.(2010)]{grise10} Gris{\'e}, F., 
Kaaret, P., Feng, H., Kajava, J.~J.~E., 
\& Farrell, S.~A.\ 2010, \apjl, 724, L148 

\bibitem[Hill et al.(2004)]{hill04} Hill, J.~E., Burrows, 
D.~N., Nousek, J.~A., et al.\ 2004, \procspie, 5165, 217 

\bibitem[Isobe et al.(2012)]{isobe12} Isobe, N., Kubota, A., 
Sato, H., \& Mizuno, T.\ 2012, \pasj, 64, 119 

\bibitem[Kalberla et al.(2005)]{kalberla05} Kalberla, P.~M.~W., Burton, W.~B., 
Hartmann, D., et al.\ 2005, \aap, 440, 775 

\bibitem[King(2009)]{king09} King, A.~R.\ 2009, \mnras, 393, L41 

\bibitem[King et al.(2001)]{king01} King, A.~R., Davies, 
M.~B., Ward, M.~J., Fabbiano, G., \& Elvis, M.\ 2001, \apjl, 552, L109 

\bibitem[King \& Dehnen(2005)]{king01b} King, A.~R., \& Dehnen, W.\ 2005, 
\mnras, 357, 275 

\bibitem[Kong et al.(2010)]{kong10} Kong, A.~K.~H., Yang, 
Y.~J., Yen, T.-C., Feng, H., \& Kaaret, P.\ 2010, \apj, 722, 1816 

\bibitem[Kubota \& Done(2004)]{kubota04} Kubota, A., \& Done, C.\ 2004, 
\mnras, 353, 980 

\bibitem[La Parola et al.(2003)]{laparola03} La Parola, V., Damiani, F.,
Fabbiano, G., Peres, G., \ 2004, \apj, 583, 758 

\bibitem[La Parola et al.(2001)]{laparola01} La Parola, V., Peres, G., 
Fabbiano, G., Kim, D. W., Bocchino, F., \ 2001, \apj, 556, 47

\bibitem[Long et al.(1981)]{long81} Long, K.~S., Dodorico, S., 
Charles, P.~A., \& Dopita, M.~A.\ 1981, \apjl, 246, L61 

\bibitem[Madau \& Rees(2001)]{madau01} Madau, P., \& Rees, M.~J.\ 2001, 
\apjl, 551, L27 

\bibitem[Makishima et al.(2000)]{makishima00} Makishima, K., 
Kubota, A., Mizuno, T., et al.\ 2000, \apj, 535, 632 

\bibitem[Middleton et al.(2015)]{middleton15} Middleton, M.~J., 
Heil, L., Pintore, F., Walton, D.~J., 
\& Roberts, T.~P.\ 2015, \mnras, 447, 3243 

\bibitem[Middleton et al.(2012)]{middleton12} Middleton, M.~J., 
Sutton, A.~D., Roberts, T.~P., Jackson, F.~E., 
\& Done, C.\ 2012, \mnras, 420, 2969 

\bibitem[Middleton et al.(2011)]{middleton11} Middleton, M.~J., 
Sutton, A.~D., \& Roberts, T.~P.\ 2011, \mnras, 417, 464 

\bibitem[Miller \& Hamilton(2002)]{miller02} Miller, M.~C., \& Hamilton, D.~P.\ 2002, \mnras, 330, 232 

\bibitem[Moretti et al.(2005)]{moretti05} Moretti, A., Campana, 
S., Mineo, T., et al.\ 2005, \procspie, 5898, 360 

\bibitem[Ohsuga et al.(2009)]{ohsuga09} Ohsuga, K., Mineshige, 
S., Mori, M., \& Kato, Y.\ 2009, \pasj, 61, L7 

\bibitem[Parmar et al.(2001)]{parmar01} Parmar, A. N., Sidoli, L.,
Oosterbroek, T., Charles, P. A., Dubus, G., Guainazzi, M., Hakala, P., 
Pietsch, W., Trinchieri, G., \ 2001, \aap, 368, 420

\bibitem[Pintore et al.(2014)]{pintore14} Pintore, F., Zampieri, 
L., Wolter, A., \& Belloni, T.\ 2014, \mnras, 439, 3461 

\bibitem[Poutanen et al.(2007)]{poutanen07} Poutanen, J., 
Lipunova, G., Fabrika, S., Butkevich, A.~G., 
\& Abolmasov, P.\ 2007, \mnras, 377, 1187 

\bibitem[Shakura \& Sunyaev(1973)]{shakura73} Shakura, N.~I., \& Sunyaev, R.~A.\ 
1973, \aap, 24, 337 

\bibitem[Soria et al.(2015)]{soria15} Soria, R., Kuntz, K.~D., 
Long, K.~S., et al.\ 2015, \apj, 799, 140 

\bibitem[Soria(2011)]{soria11} Soria, R.\ 2011, Astronomische 
Nachrichten, 332, 330 

\bibitem[Stobbart et al.(2006)]{stobbart06} Stobbart, A.-M., 
Roberts, T.~P., \& Warwick, R.~S.\ 2006, \mnras, 370, 25 

\bibitem[Sutton et al.(2012)]{sutton12} Sutton, A.~D., Roberts, 
T.~P., Walton, D.~J., Gladstone, J.~C., 
\& Scott, A.~E.\ 2012, \mnras, 423, 1154 

\bibitem[Sutton et al.(2013)]{sutton13} Sutton, A.~D., Roberts, 
T.~P., \& Middleton, M.~J.\ 2013, \mnras, 435, 1758 

\bibitem[Swartz et al.(2011)]{swartz11} Swartz, D.~A., Soria, 
R., Tennant, A.~F., \& Yukita, M.\ 2011, \apj, 741, 49 

\bibitem[Trinchieri et al.(1988)]{trinchieri88} Trinchieri, G., 
Fabbiano, G., \& Peres, G.\ 1988, \apj, 325, 531 

\bibitem[Vierdayanti et al.(2010)]{vierdayanti10} Vierdayanti, K., 
Done, C., Roberts, T.~P., \& Mineshige, S.\ 2010, \mnras, 403, 1206 

\bibitem[Ueda et al.(2009)]{ueda09} Ueda, Y., Yamaoka, K., 
\& Remillard, R.\ 2009, \apj, 695, 888 

\bibitem[Walton et al.(2014)]{walton14} Walton, D.~J., Harrison, 
F.~A., Grefenstette, B.~W., et al.\ 2014, \apj, 793, 21 

\bibitem[Watarai et al.(2000)]{watarai00} Watarai, K.-y., Fukue, 
J., Takeuchi, M., Mineshige, S., \ 2000, \pasj, 52, 133 


\bibitem[Weng et al.(2009)]{weng09} Weng, S.-S., Wang, J.-X., 
Gu, W.-M., \& Lu, J.-F.\ 2009, \pasj, 61, 1287 

\bibitem[Winter et al.(2006)]{winter06} Winter, L.~M., 
Mushotzky, R.~F., \& Reynolds, C.~S.\ 2006, \apj, 649, 730 

\end{thebibliography}

{}

\end{document}